\documentstyle[epsf,mystylr]{article}
\epsfverbosetrue

\font\tenrm=cmr10
\font\tenit=cmti10
\font\elevenbf=cmbx10 scaled\magstep 1
\font\elevenrm=cmr10 scaled\magstep 1
\font\elevenit=cmti10 scaled\magstep 1

\renewenvironment{thebibliography}[1]
 { \elevenrm
   \begin{list}{\arabic{enumi}.}
    {\usecounter{enumi} \setlength{\parsep}{0pt}
     \setlength{\itemsep}{3pt} \settowidth{\labelwidth}{#1.}
     \sloppy
    }}{\end{list}}

\newcommand{\tb}{\tan \beta}
\newcommand{\wti}{\widetilde}

\newcommand{\bsgamma}{b\to s \gamma}

\newcommand{\thb}{t\to H^+b}
\newcommand{\tstopneu}{t\to {{\wti u}_1} {{\wti \chi}^0_1}}
\newcommand{\nn}{\nonumber}
\def\npb#1#2#3{    {\elevenit Nucl. Phys. }{\elevenbf B#1} (19#2) #3}
\def\plb#1#2#3{    {\elevenit Phys. Lett. }{\elevenbf B#1} (19#2) #3}
\def\prd#1#2#3{    {\elevenit Phys. Rev. }{\elevenbf D#1} (19#2) #3}
\def\prep#1#2#3{   {\elevenit Phys. Rep. }{\elevenbf #1} (19#2) #3}
\def\prl#1#2#3{    {\elevenit Phys. Rev. Lett. }{\elevenbf #1} (19#2) #3}
\def\ptp#1#2#3{    {\elevenit Prog. Theor. Phys. }{\elevenbf #1} (19#2) #3}
\def\rmp#1#2#3{    {\elevenit Rev. Mod. Phys. }{\elevenbf #1} (19#2) #3}
\def\zpc#1#2#3{    {\elevenit Z. Phys. }{\elevenbf C#1} (19#2) #3}
\def\mpla#1#2#3{   {\elevenit Mod. Phys. Lett. }{\elevenbf A#1} (19#2) #3}
\def\nc#1#2#3{     {\elevenit Nuovo Cim. }{\elevenbf #1} (19#2) #3}
\def\yf#1#2#3{     {\elevenit Yad. Fiz. }{\elevenbf #1} (19#2) #3}
\def\sjnp#1#2#3{   {\elevenit Sov. J. Nucl. Phys. }{\elevenbf #1} (19#2) #3}
\def\jetp#1#2#3{   {\elevenit Sov. Phys. }{JETP }{\elevenbf #1} (19#2) #3}
\def\jetpl#1#2#3{  {\elevenit JETP Lett. }{\elevenbf #1} (19#2) #3}
\begin{document}
\begin{titlepage}
\noindent
DESY 93-099     \hfill         ISSN 0418-9833     \\
October 93       \hfill                             \\[7ex]
\begin{center}
{\bf SUPERSYMMETRIC TOP QUARK DECAYS
\footnote{Contribution to the Workshop on $500\,$GeV $e^+e^-$
Linear Collider, Hamburg, April 1993.     }    }           \\[11ex]

{\bf    Francesca M.\ Borzumati     }                      \\[1ex]
{\it II.\ Institut f\"ur Theoretische Physik
\footnote{ Supported by the Bundesministerium f\"ur Forschung und
 Technologie, 05 5HH 91P(8), Bonn, FRG.   } }           \\
{\it Universit\"at Hamburg, 22761 Hamburg, Germany}   \\[12ex]
\parbox{13.5cm}
{\begin{center} ABSTRACT \end{center}
\vspace*{-1mm}
{
\noindent
The supersymmetric decays of the top quark into charged Higgs plus
bottom, $\thb$, and into the supersymmetric partner of the top
(${\wti u}_1$) plus the lightest neutralino (${\wti \chi}_1^0$),
$\tstopneu$, are discussed within the framework of the Minimal
Supersymmetric Standard Model with radiatively induced breaking of
the gauge group $SU(2)\times U(1)$. The possibility of detecting
these decays at present, i.e. given the available bounds on
supersymmetric
parameters, is compared with the situation a Next $e^+e^-$ Linear
Collider would face if supersymmetric particles were still
undiscovered at LEP~II. The indirect implications for $\thb$ and
$\tstopneu$ of a measurement of the bottom quark decay $\bsgamma$ at
the Standard Model level are taken into account.    }}
\vfill
\end{center}
\end{titlepage}
\newpage
\begin{center}
\vglue 0.6cm
{{\elevenbf  SUPERSYMMETRIC DECAYS OF THE TOP QUARK   \\}
%
\vglue 1.0cm
{\tenrm FRANCESCA M.~BORZUMATI \\}
\baselineskip=13pt
{\tenit  II.\ Institut f\"ur Theoretische Physik
\footnote{Supported by the Bundesministerium f\"ur Forschung und
 Technologie, 05 5HH 91P(8), Bonn, FRG.   },
            Universit\"at Hamburg \\}
\baselineskip=12pt
{\tenit 22761 Hamburg, Germany \\} }
\vglue 0.8cm
{\tenrm ABSTRACT}

\end{center}

\vglue 0.3cm
{\rightskip=3pc
 \leftskip=3pc
 \tenrm\baselineskip=12pt
 \noindent
The supersymmetric decays of the top quark into charged Higgs plus
bottom, $\thb$, and into the supersymmetric partner of the top
(${\wti u}_1$) plus the lightest neutralino (${\wti \chi}_1^0$),
$\tstopneu$, are discussed within the framework of the Minimal
Supersymmetric Standard Model with radiatively induced breaking of
the gauge group $SU(2)\times U(1)$. The possibility of detecting
these decays at present, i.e. given the available bounds on
supersymmetric
parameters, is compared with the situation a Next $e^+e^-$ Linear
Collider would face if supersymmetric particles were still
undiscovered at LEP~II. The indirect implications for $\thb$ and
$\tstopneu$ of a measurement of the bottom quark decay $\bsgamma$ at
the Standard Model level are taken into account.  }

\vglue 0.6cm
{\elevenbf\noindent 1. Introduction}
\vglue 0.4cm
\baselineskip=14pt
\elevenrm

\looseness=-1
The two supersymmetric decays of the top quark into charged Higgs
plus bottom, $\thb$, and into the supersymmetric partner of the
top (${\wti u}_1$) plus the lightest neutralino (${\wti \chi}_1^0$),
$\tstopneu$, as well as the subsequent decays of ${\wti u}_1$
have been extensively discussed in the past;
see~\cite{REW}-\cite{PAST2}. No investigation, however, has been
performed to
find out whether the correct ranges of masses needed for these
decays are actually present in realistic supersymmetric models.

\looseness=-1
The aim of this paper is to show the results of such an
investigation and to discuss the present and future prospects for
these two decays. The framework considered is the Minimal
Supersymmetric Standard Model (MSSM) with breaking of the gauge group
$SU(2)\times U(1)$ induced by renormalization effects of the
mass parameters in the tree-level potential of the neutral Higgs
sector of this model. The embedding in a generic grand-unified
scheme is also considered. Details regarding the procedure
followed and the approximations made in this search are given
in~\cite{MEAGAIN}. I remind only that three new parameters
$(m,M,\tb)$ in addition to those present in the Standard Model are
needed to completely specify all masses and couplings in this
model.

\looseness=-1
The present experimental situation is simulated in this search by
imposing lower
bounds on the masses of supersymmetric particles which roughly
match the limits coming from LEP~I and the TEVATRON. To be specific,
the cuts applied to the masses of gluinos $\wti g$,
charginos ${\wti \chi}^-$ (among which the
lightest is conventionally denoted as ${{\wti \chi_2}^-}$),
neutralinos ${\wti \chi}^0$, charged and neutral sleptons
$\wti l$,~$\wti \nu$, up- and down-squarks $\wti u$,~$\wti d$, and
neutral Higgses (with $h_2^0$ the lightest of the two CP-even
states) are:
\bea
                          &      &
m_{\wti g}            \ \     >   120\,{\rm GeV},   \quad\quad
m_{{\wti d_1},{\wti u_2}}     >   100\,{\rm GeV},   \quad\quad
m_{\wti u_1}                  >  \ 45\,{\rm GeV},   \qquad\quad  \nn \\
& {\rm SET \quad I\phantom{I}}:\quad   &
m_{\wti\chi^-_2}              >\ \ 45\,{\rm GeV},   \quad\quad
m_{\wti\nu_1}         \  \  \,>\,\ 45\,{\rm GeV},   \quad\quad
m_{\wti l_1}                \,>\   45\,{\rm GeV},   \qquad\quad  \nn \\
                          &      &
m_{h_2^0}                   \,>\ \ 30\,{\rm GeV},   \quad\quad
m_{\wti\chi^0_1}      \  \  \,>\,\ 20\,{\rm GeV}.   \qquad\quad
\label{setone}
\eea

\begin{figure}[t]
\epsfxsize=7.6 cm
\leavevmode
\caption[f5]{{Allowed values of $m_{{H^-}}$ for the decay
 $\thb$   }}
\label{beforebsg}
\end{figure}

Except for squarks and gluinos masses, the lower bounds
in~(\ref{beforebsg}) are roughly the limits coming from LEP. The
lower bounds imposed on squarks and gluinos masses are lower than
the values of $126$ and $141\,$GeV respectively excluded by CDF
when no cascade decays are considered. Cascade decays become
possible when squarks and gluinos are heavy enough to decay into
charginos and neutralinos before decaying into the lightest
supersymmetric particle. Since such a circumstance is very often
encountered in the MSSM, the abovementioned CDF limits are
reduced in the search here performed. Similarly, it is reasonable
to assume that the bounds applicable to squarks and gluinos masses
in the MSSM will not increase very much in the near future. Thus,
the situation which
the Next Linear $e^+e^-$ Collider could have to face if
supersymmetric particles were undiscovered at LEP~II can be
mimicked by imposing the following (conservative) lower bounds:
\bea
                          &      &
m_{\wti g}            \ \     >   140\,{\rm GeV},   \quad\quad
m_{{\wti d_1},{\wti u_2}}     >   120\,{\rm GeV},   \quad\quad
m_{\wti u_1}                  >  \ 80\,{\rm GeV},   \qquad\quad  \nn \\
& {\rm SET \quad II}:\quad       &
m_{\wti\chi^-_2}              >\ \ 80\,{\rm GeV},   \quad\quad
m_{\wti\nu_1}        \  \   \,>\,\ 80\,{\rm GeV},   \quad\quad
m_{\wti l_1}                \,>\   80\,{\rm GeV},   \qquad\quad  \nn \\
                          &      &
m_{h_2^0}                   \,>\ \ 70\,{\rm GeV},   \quad\quad
m_{\wti\chi^0_1}     \  \   \,>\,\ 40\,{\rm GeV}.   \qquad\quad
\label{settwo}
\eea
In the following, I shall refer to these two choices of
bounds~(\ref{setone}) and~(\ref{settwo}) as SET~I and SET~II. In
both cases, for the limits on the lightest eigenvalue of the
up-squark mass matrix ($m_{\wti u_1}$) one relies on the
LEP~I and LEP~II searches instead than the searches at the TEVATRON,
where the decay ${\wti u_1} \to t + {\wti \chi^0_1}$ is forbidden or
disfavoured. Finally, the limits on the mass of the lightest
neutralino are in the MSSM induced by the limits on the lightest
chargino mass.

The results shown in this paper are limited to
$m_t=150\,{\rm GeV}$, but similar features are obtained for
different values of $m_t$.

\vglue 0.6cm
{\elevenbf\noindent 2. Decay $\thb$}
\vglue 0.4cm
\looseness=-1
I have performed a systematic study of the $(m,M)$ supersymmetric
parameter space in the range $0\leq m \leq 500\,{\rm GeV}$,
$-250 \leq M \leq 250\,{\rm GeV}$ for several values of $\tb$,
between $3$ and $35$ (for the rationale behind this
choice of values see~\protect\cite{MEAGAIN}).

\begin{figure}[t]
\epsfxsize=7.6 cm
\leavevmode
\caption[f5]{{$\tb$ dependence of the ratio $R_H$}}
\label{ratio}
\end{figure}

The obtained masses $m_{H^-}$ for which the decay $\thb$
is kinematically allowed are shown in fig.~\ref{beforebsg}.
\hangover
When SET~I of lower bounds is imposed, masses as small as
$90\,{\rm GeV}$ are obtained for the largest values of
$\tb$ considered (from $28$ to $35$). The available phase space
(the running mass of the b-quark $m_b(M_Z)$ is here
$\sim 3.6\,{\rm GeV}$) shrinks rapidly when $\tb $ decreases,
reaching a minimum around $\tb =9$, for which the smallest
$m_{H^-}$ obtained is $\sim 120\,{\rm GeV}$. The minimum allowed
value of $m_{H^-}$ tends to increase again up to
$m_{H^-} \sim 100\,{\rm GeV}$ when $\tb$ is reduced from
$9$ to $3$.

\begin{figure}[t]
\epsfxsize=7.6 cm
\leavevmode
\caption[f5]{{$R_H$ for the points in fig.~\ref{beforebsg},~SET~I}}
\label{wdhb}
\end{figure}

Not visible in this figure is the fact that the lower values of
masses obtained for $\tb\!=\!3$ are less ``probable'' than for
large $\tb$, since obtained in much smaller regions of the same
$(m,M)$ parameter space considered; for details see~\cite{MEAGAIN}.
\hangover
The results described here dif\-fer from those presented
in~\cite{OLECHPOK}. The di\-sa\-greement may be due to
dif\-fe\-ren\-ces in
the assumptions underlying the two calculations, al\-though the
precise reason is not yet clear.
\hangover
A strong dependence on $\tb $ is present also in the expression for
the width $\Gamma (\thb)$. The ratio
$R_H\equiv \Gamma(\thb)/\Gamma(t \to W^+ b)$ is plotted in
fig.~\ref{ratio} versus $\tb$, for $m_t=150\,$GeV and three different
values of $m_{H^-}$. This figure points clearly to the intermediate
values of $\tb$ as to the disfavored ones.

The overall $(m_{H^-},\tb)$ dependence of this ratio for the phase
space of fig.~\ref{beforebsg}, SET~I, is shown in fig.~\ref{wdhb}.
The value $\tb= 3$ yields a maximum ratio $R_H$ of about $5\%$;
intermediate values, $3\!<\!\tb\!<\!15$, are doubly penalized
by the intrinsic drop in the rate shown in fig.~\ref{ratio} and the
fact that the masses $m_{H^-}$ obtained in these cases are heavier;
values of $\tb$ greater than $25$ can give ratios $R_H$ as big as
$20\!-\!30\%$. The subsequent decay of $H^-$ into the pair
$\tau,\nu_\tau$, with a rate practically equal to one (for the values
of $\tb$ considered here) contributes to making the prospects for the
detection of this decay mode rather
optimistic; for a discussion on this issue see~\protect\cite{VENTURI}.

The tight relations among the mass parameters in the MSSM, however,
are such that an increase of the lower bounds of supersymmetric
particles up to the values of SET~II drastically reduces the phase
space available for this decay, as shown in the second plot of
fig.~\ref{beforebsg}. The expected rates, which can be read
with some effort from fig.~\ref{wdhb}, are still far from small.

\begin{figure}[tb]
\epsfxsize=7.6 cm
\leavevmode
\caption[f5]{{What remains of fig.~\ref{beforebsg} after
    imposing the constraints coming from $\bsgamma$   }}
\label{afterbsg}
\end{figure}

It has been recently argued~\cite{HEWBAGG} that the the channel
$\thb$ could be closed by a measurement of the bottom quark decay
$\bsgamma$ compatible with
the SM prediction. A careful analysis of the constraints induced by
this decay on the supersymmetric parameter space, however, brings
in different results. The ranges of masses $m_{H^-}$ remaining
after imposing that the values of $BR(\bsgamma)$ in the MSSM be
limited to the interval $2.9\!-\!4.8\times 10^{-4}$ (the range of
allowed values of $BR(\bsgamma)$ within the SM for
$m_t=150\,$GeV~\cite{MEAGAIN}) are shown in fig.~\ref{afterbsg} for
the two sets of lower bounds SET~I and SET~II. Although reduced
(and for some values of $\tb$ even drastically reduced), the
regions of supersymmetric parameter space where the decay $\thb$ is
kinematically allowed are not closed by a measurement
of $\bsgamma$. The probability of the top quark decay into
the supersymmetric charged Higgs, however, becomes practically
negligible if the lower bounds SET~II are implemented, see second
plot in fig.~\ref{afterbsg}.

\vglue 0.5cm
{\elevenbf \noindent 3. Decay $\tstopneu$}
\vglue 0.4cm

The presence of large left-right entries in the up-squark mass
matrix is such to allow one light mass eigenvalue. The corresponding
eigenstate ${\wti u}_1$, or stop, much lighter than the remaining
squarks, can be in certain region of the supersymmetric parameter
space even lighter than the top quark. Moreover, also the mass of
the lightest neutralino, ${{\wti \chi}_1}^0$, can be rather light
within the MSSM.

\begin{figure}[t]
\epsfxsize=9.68 cm
\leavevmode
\caption[f5]{{Phase space for the decay $\tstopneu$, SET~II}}
\label{stneu}
\end{figure}
\looseness=-1
An investigation of the 2-dimensional parameter space $(m,M)$ for
six values of $\tb$,~i.e.~$3,9,15,20,25$, $30$, leads to the results
in fig.~\ref{stneu} for the allowed phase space of the decay
$\tstopneu$. The bounds SET~I are implemented in this figure.
\hangover
As in the case of $\thb$, the region of parameter space studied
is limited to the range
$0\leq m \leq 500\,{\rm GeV}$,~$-250 \leq M \leq 250\,{\rm GeV}$,
except for $\tb\!=\!30$. In this case, in fact, a wider region has
been explored since viable masses
$m_{{\wti u}_1}$,~$m_{{\wti \chi_1}^0}$ are obtained for $m$ and
$\vert M \vert$ as big as $700\,{\rm GeV}$ and $350\,{\rm GeV}$,
respectively.
\hangover
Also for this decay,
the size of the allowed phase space
decreases for increasing values of $\tb$, reaches a minimum, and
starts increasing again after $\tb=15$.

\begin{figure}[t]
\epsfxsize=7.6 cm
\leavevmode
\caption[f5]{{Rates for the decay $\tstopneu$}}
\label{gone}
\end{figure}
The dependence on $\tb$ of the width $\Gamma(\tstopneu)$ is
less pronounced than for the decay $\thb$. The ratios
$R_{{\wti u}_1} \equiv \Gamma(\tstopneu)/\Gamma(t \to W^+ b)$
obtained for all the values of $\tb$ considered in this case,
are plotted in fig.~\ref{gone}: for the smallest values of
$m_{{\wti u}_1}$ obtained, $R_{{\wti u}_1}$ can be as high as
$ 10\%$.
\hangover
The produced ${\wti u}_1$ can then decay as
${\wti u}_1 \to {\wti \chi}_2^+ b$, with ${\wti \chi}_2^+$ the
lightest chargino, if $m_{{\wti \chi}_2^+} < m_{{\wti u}_1}$.
This is the case for all the points in fig.~\ref{stneu} located
beneath the solid lines. The corresponding regions where the
relation $m_{{\wti \chi}_2^+} < m_{{\wti u}_1}$ is satisfied,
practically coincide, except for $\tb=3,30$, with the complete
regions of phase space obtained.

\begin{figure}[t]
\epsfxsize=7.6 cm
\leavevmode
\caption[f5]{{Points of fig.~\ref{gone} plotted versus $m_{H^-}$}}
\label{gsth}
\end{figure}

For the remaining regions where
${\wti \chi_2^+}$ is heavier than ${{\wti u}_1}$,
the alternatives are, in principle, the three-body decays
mediated by virtual charginos, i.e.
${\wti u}_1 \to b l^+ {\wti \nu}_1$,~$b {\wti l}_1^+ \nu$, and
${\wti u}_1 \to b W^+ {\wti \chi}^0_1$ mediated by a virtual
chargino and/or a down-squark. Given the values of $m_{{\wti u}_1}$,
$m_{{\wti \chi_1}^0}$ in the points above the solid lines in
fig.~\ref{stneu}, however, it is clear that this last possibility
can only occur with off-shell $W$-bosons.
\hangover
Mo\-reo\-ver, an explicit check of the values of masses which
sleptons acquire in the regions of phase space whe\-re
$m_{{\wti \chi_2}^+} > m_{{\wti u}_1}$, leads to the conclusion that
also ${\wti u}_1 \to b l^+ {\wti \nu}_1$, $b {\wti l}_1^+ \nu$ occur
as four-body decays, with virtual sleptons producing leptons plus
neutralinos. The pro\-spects of sizeable decay rates for stop lighter
than charginos look, therefore, rather grim.

\begin{figure}[t]
\epsfxsize=9.68 cm
\leavevmode
\caption[f5]{{Phase space for the decay $\tstopneu$, SET~II}}
\label{stn2}
\end{figure}

\looseness=-1
The MSSM, however, provi\-des another interesting possibility,
i.e. the two-body Flavour Changing Neutral Current decay mode
$ {\wti u}_1 \to c {\wti \chi_1}^0$. This is mediated by the
coupling ${\wti \chi}^0-u_i-{\wti u_j}$ ($i\neq j$) induced
through renormalization effects by the soft
super\-symmetry-\-brea\-king
\newline
terms. In spite of the small coupling,
this decay mode is clearly winning when compared to the four-body
decay channels, which suffer from severe phase space
suppressions~\cite{JAP}.
\hangover
It is interesting to observe that the two decays $\thb$,~$\tstopneu$,
are sensitive to different regions of the supersymmetric parameter
space. High rates for $\tstopneu$ can be obtained in regions were
$m_{H^-}$ is already far too big for $\thb$ to be kinematically
allowed. This is explicitly shown in fig.~\ref{gsth} where the points
of fig.~\ref{gone} are plotted versus $m_{H^-}$ instead than versus
$m_{{\wti u}_1}$.

\begin{figure}[t]
\epsfxsize=7.6 cm
\leavevmode
\caption[f5]{{Rates obtained for limits SET~II}}
\label{gsth2}
\end{figure}

Therefore, one may think that in the case of heavy charged Higgses,
the decay $\tstopneu$ would be the dominant top-decay-mode, if all
the other supersymmetric masses could remain unchanged. An increase
in $m_{{H^-}}$, however, brings up with itself also other
masses. As it can be seen in fig.~\ref{stn2}, the reduction of
the available phase space when the limits SET~II are applied, is even
more severe than the one suffered by $\thb$. Only two small
regions of points remain for $\tb=3$ and $30$. A comparison with
fig.~\ref{stneu} shows that both type of decays for the stop squark,
i.e. ${\wti u}_1 \to {\wti \chi}_2^+ b$ and
$ {\wti u}_1 \to c {\wti \chi_1}^0$, are possible.
\hangover
The ratio $R_{{\wti u}_1}$ can still reach in these two regions
values as high as $5-6\%$, in points where
$m_{{H^-}}>>150\,{\rm GeV}$. The scarcity of points of
fig.~\ref{gsth2}, however, shows how unlikely this decay mode
becomes when the lower bounds on the supersymmetric particles are
increased.

Moreover, very few points of the phase space shown in
fig.~\ref{stneu}, for SET~I, remain when the constraints due to
a measurement of the decay $\bsgamma$ are implemented. They
are all found for values of $m_{{\wti u}_1} \gtap 80\,{\rm GeV}$.
The possibility of detecting $\tstopneu$ seems to be completely
closed if these contraints are implemented on the phase space
obtained when the limits SET~II are imposed.

\vglue 0.5cm
{\elevenbf \noindent 5. Acknowledgements \hfil}
\vglue 0.3cm
It is a pleasure to thank the organizers of this conference and the
colleagues of the Top Working Group. Discussions with F.~Hautmann
are acknowledged.
\vglue 0.5cm
{\elevenbf\noindent 6. References \hfil}
\vglue 0.3cm

%
\end{document}